\documentclass[aps,prl,twocolumn,groupedaddress]{revtex4-1}

\usepackage[dvips]{graphicx}

\begin{document}


\title{Role-based similarity in directed networks}


\author{Kathryn Cooper}

\author{Mauricio Barahona}
\affiliation{Department of Bioengineering \& Institute for Mathematical Sciences, Imperial College London, South Kensington Campus, London SW7 2AZ, United Kingdom}


\date{\today}

\begin{abstract}
 The widespread relevance of increasingly complex networks requires methods to extract meaningful coarse-grained representations of such systems. 
 For undirected graphs, standard community detection methods use criteria largely based on density of connections to provide such representations.   
We propose a method for grouping nodes in directed networks based on the role of the nodes in the network, understood in terms of patterns of incoming and outgoing flows. The role groupings are obtained through the clustering of a similarity matrix, formed by the distances between feature vectors that contain the number of in and out paths of all lengths for each node. Hence nodes operating in a similar flow environment are grouped together although they may not themselves be densely connected.  Our method, which includes a scale factor that reveals robust groupings based on increasingly global structure, provides an alternative criterion to uncover structure in networks where there is an implicit flow transfer in the system. We illustrate its application to a variety of data from ecology, world trade and cellular metabolism. 
\end{abstract}


\maketitle


The recent surge of interest in the study of complex networks spans diverse disciplines, from physics and computer science to biology and the social sciences. Classic examples include the Internet, protein interaction networks, food webs or social groupings, among many others.~\cite{Girvan:2002p46,Strogatz01,Newman:2003p39, Boccaletti2006175}.  A network is a collection of nodes connected by edges that represent interactions. In many instances, the edges have an associated direction or weight but the vast majority of research to-date has focused upon unweighted and undirected graphs.    Network representations have the advantage that they capture naturally properties at the system level starting from individual constituents. However, with the growth of computational capability and high-throughput technologies, network representations quickly become so complex as to lack intelligibility. 

A key challenge in this area is the development of methods to obtain simplified reduced representations of complex networks in terms of subgraphs or communities, i.e., meaningful groupings of nodes that are significantly related. For instance, nodes are likely to belong together if they are part of a tightly-knit group with many connections within the group and fewer to external nodes. The flurry of research on clustering of networks and community detection~\cite{Fortunato201075} has led to the rediscovery of classic results in graph partitioning, and to the development of new measures such as modularity~\cite{Newman:2006p26} and various spectral algorithmic procedures~\cite{Ng01onspectral, Shi:2000p825}. Most methods have focused on undirected networks, where it is natural to consider structural metrics based on the density of intra- and inter-community edges. However, there is a large class of networks where the directionality of the edges is essential and where an analysis based on undirected graphs risks missing key properties of the system. Examples include social networks, food webs, the world wide web and systems involving causality, such as metabolic and genetic networks. Only recently, extensions of notions of modularity for directed graphs have been proposed~\cite{LeichtNewmanDirected, Kim2010} as well as other measures based on diffusion dynamics that can be applied to both directed and undirected graphs~\cite{Rosvall01052007, Delvenne20072010}.

Here, we introduce an alternative measure for the grouping of nodes in directed networks. Given that the defining characteristic of directed graphs is the implicit existence of flows, we propose to group nodes according to their {\it role} in the network, defined in terms of the overall pattern of incoming and outgoing flows.  Essentially, the profile of paths for each node is a vector that is computed from the powers of the adjacency matrix weighted with a scale parameter to yield a similarity matrix, defined by the distances between such node vectors. This matrix is then clustered to find groupings of nodes with similar profiles of reachability flows at all lengths. For instance, in our analysis, all nodes that are sources are found to be similar to each other, while sinks are grouped together.  In between these extremes, nodes are grouped according to a quantitative measure that reflects the mixture of `hub' vs.\ `authority' characteristics of each node with respect to all paths in the graph. Our definition is inspired by a vast array of literature from the social sciences, dealing with structural and regular equivalence~\cite{deNooy:2005p3733, Borgatti:1993p2749, Freeman19781979215, Reichardt2007}, and from computer science, where alternative algorithmic measures of similarity have been considered~\cite{Blondel:2004p2427,SimRank, Kleinberg:1999, Leicht:2005}.

Our methodology can be used to unveil groups distinct to those found by community detection algorithms based on density of connections. Indeed, nodes that play similar roles may be only weakly connected. For instance, in a food-web, two predators are not likely to be linked directly although both perform the same function and would be canonically grouped within the same trophic level. 
Hence, role similarity can uncover a coarse-grained functional representation for networks where the dominating feature is the transfer of an underlying quantity (e.g., information, energy, matter, etc).  This role-based representation is relevant in fields such as ecology, economics, social sciences and cellular metabolism, where it can aid in the assignment of a putative function to uncharacterised nodes and in establishing functional relations between seemingly distant network elements.
\begin{figure}[htbp]
\includegraphics[]{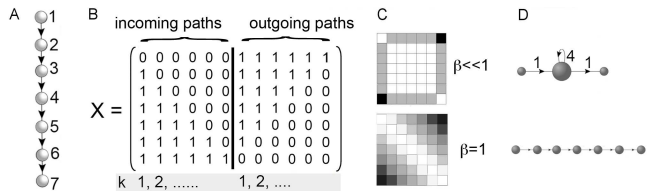}
\caption{Role clustering for the directed path graph in \textit{(A)}.  The construction of the flow matrix $X$ (shown in \textit{(B)} for $\beta=1$) is followed by the construction of the similarity matrix $Y$ (shown in grayscale in \textit{(C)} for $\beta \ll 1$ and $\beta=1$). 
The resulting role groupings of the nodes are shown in \textit{(D)}.\label{fig:fig1}}
\end{figure}

The measure is defined as follows. Consider a directed graph with $N$ nodes and adjacency matrix $A$, which is in general asymmetric.
The number of outgoing paths of length $k$ for node $i$ is given by the $i$-th coordinate of the vector 
$[A^{k}\mathbf{1}]$, where $\mathbf{1}$ is the $N \times 1$ vector of ones.  Similarly, the number of incoming paths of length $k$ for node $i$ is: 
$[{A^T}^k\mathbf{1}]_i$. Note that the case $k=1$ corresponds to the out-degree and in-degree which, from this perspective, represent the number of paths of length one originating or terminating at the node.  

We now construct a matrix that compiles the incoming and outgoing paths of all lengths up to $k_{max}$ by appending the column vectors indexed by path length and scaled by the factors $\beta^k$: 
$$
X = \left [  \begin{array}{c}  \mathbf{x}_1  \\ \vdots \\ \mathbf{x}_N  \end{array} \right]  \equiv  [\,\,\underbrace{ \ldots \,\, (\beta A^{T})^k\mathbf{1} \,\, \ldots }_{k_{max}} | \underbrace{\ldots \,\, (\beta A)^{k}\mathbf{1} \,\, \ldots}_{k_{max}} \,\, ].
$$
Here, $\beta=\alpha/\lambda_1$, with $\lambda_1$ the largest eigenvalue of the adjacency matrix and $0 \leq \alpha \leq 1$. The parameter $\alpha$ is a \textit{scale factor} that allows us to tune the weight of the local environment (short paths) relative to the global network structure (long paths). The presence of the factors $\beta^k$ ensures the convergence of the sequence of the columns due to the asymptotic limit $\lim_{k \to \infty} \frac{||A^{k+1}||}{||A^k||} \rightarrow \lambda_1$.~\cite{Leicht:2005}

Each \textit{row vector} of $X$ contains the flow profile of a node in terms of the scaled number of incoming and outgoing paths of all lengths starting and ending at that node (see Fig. 1).  Our criterion to group nodes together is that they have similar flow profiles. This can be quantified via a distance between the vectors  $\mathbf{x}_i$. A simple choice of metric is the cosine distance, which leads to the symmetric \textit{similarity matrix} $Y$ defined by:
\begin{equation}
Y_{ij}= \frac{\mathbf{x}_i  \mathbf{x}^T_j}{||\mathbf{x}_i|| \, ||\mathbf{x}_j||},
\end{equation}
where element $Y_{ij}$ provides a normalized measure of the closeness of the flow profiles of nodes $i$ and $j$. Dissimilar flow profiles have a similarity value close to zero, while alike flow profiles lead to a similarity close to one.  The groupings of nodes are obtained from the clustering of this similarity matrix:  nodes in the same cluster have similar flow profiles and can be considered to play a similar role in terms of the flow in the directed graph. It is important to remark that the clustering of the similarity matrix can be performed with any of a variety of methods available for weighted symmetric graphs. In what follows, we have chosen a spectral algorithm based on a Multiple Normalized Cut~\cite{Shi:2000p825, Azran:2006p1031} together with Gaussian preprocessing  of the weights. However, the results do not depend heavily on the choice of clustering algorithm. 

Our procedure is illustrated in Figure~\ref{fig:fig1} through the simple example of a path graph. We scan the groupings as a function of the scale factor 
$\alpha$ so as to reveal role groupings based on an increasingly global flow structure.  
When $\alpha$ is small, short path lengths dominate and nodes are classified in terms of their local properties. In the limit $\alpha \to 0$, only paths of length one contribute to the clustering, which is equivalent to classifying nodes according to their in- and out-degree. In this limit, the nodes of the path graph (Fig.~\ref{fig:fig1}) are classified into three groups according to their role: input $\to$ intermediate $\to$ output, i.e., all the internal nodes are identical based on their short-scale patterns of in- and out-flows.
As $\alpha$ grows towards 1, longer paths are given increasingly more weight and the global flow structure of the network is taken into account to cluster the nodes.  For the path graph in Fig.~\ref{fig:fig1} taking into account the global structure (in this case, the presence of end nodes) means that each node is classified as having a different role. In some simplified examples, the groupings are identical at all values of the scale parameter, as in the test examples in~\cite{LeichtNewmanDirected} (not shown) in which the nodes can be distinguished based on their in- and out-degrees. Similarly, the flow example presented in~\cite{Rosvall01052007} is also reduced into a meaningful representation of two groups (not shown). In general, however, robust non trivial clusterings are found for values of $\alpha \to 1$ in more complex examples. 

\begin{figure}[htbp]
\includegraphics[]{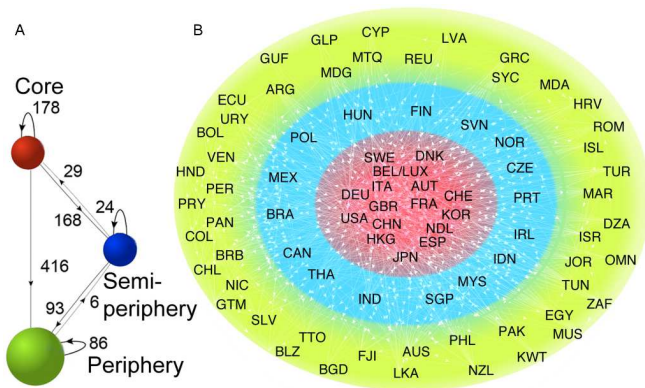}
\caption{World trade network of manufacture of metals. Our algorithm finds a robust grouping  in which each country is classified into core, semi-periphery and periphery. The color-coded reduced representation is shown in \textit{(A). } \label{fig:figWT}}
\end{figure}

We have used our method to analyze several types of networks from real data where flows are intrinsic to the system. Below, we present three examples taken from the Social Sciences, Ecology and Biochemistry.  Figure~\ref{fig:figWT} shows the role-classification in a world-trade network of manufacture of metals in 1994~\cite{deNooy:2005p3733,Smith:1991p3591}. A well-established concept in this literature is that the world economy can be broken down into a core, a semi-periphery and a periphery.  Dominant core countries tend to specialize in high-tech production requiring capital, whereas peripheral countries supply raw materials and labor intensive products.  As a consequence, there tend to be lots of connections within the core but few trade connections between members of the periphery.  Figure~\ref{fig:figWT} shows that our algorithm finds a robust classification into three groups that can be ascribed to this conceptual framework.

\begin{figure}[htbp]
\includegraphics[]{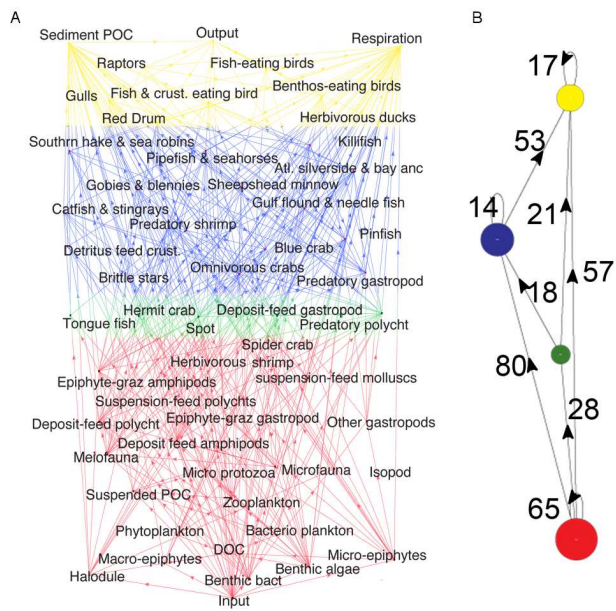}
\caption{Analysis of an ecological example, the St Mark's foodweb, showing the position of each species into role groupings akin to trophic levels \textit{(A)} with the reduced depiction \textit{(B)} indicating the flow of carbon through the network.  \label{fig:figFW}}
\end{figure}

Our second example is the food web of St Marks river in Florida~\cite{Luczkovich:2003p2761,Baird:1998p3522}. This ecological network is
underpinned by an underlying flow of carbon (i.e., assimilated matter and energy). Importantly, trophic levels are not defined by the density of internal connections but rather by their role (or position) within the flows of the network. Figure~\ref{fig:figFW} shows that the groupings produced by our algorithm detect trophic levels with the expected content.  Carbon producers such as algae and bacteria are grouped together with other basal taxa as sources in the network. Above these are small bottom-feeding fish such as Spot and Tongue fish, as well as some benthic invertebrates.  One more level up are most fish, along with some predatory invertebrates such as shrimp and omnivorous crabs.  The top level consists of all birds, large predatory fish and other sinks of the system. 

\begin{figure}[htbp]
\includegraphics[]{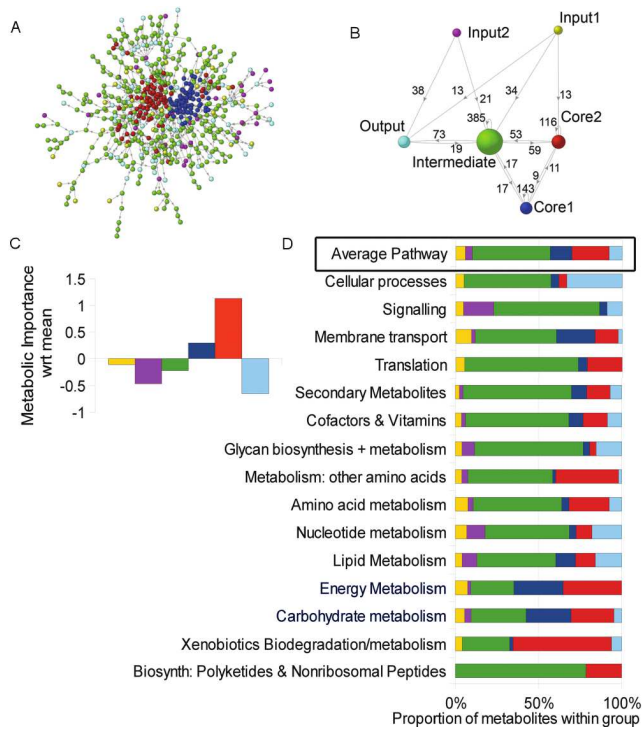}
\caption{Role-based clustering of the metabolic network of E.\ coli (largest connected component with $N=563$ nodes): force-based Kamada-Kawai layout of the network \textit{(A)} and reduced representation \textit{(B)} showing the position of each role grouping.  \textit{(C)} The metabolic importance of both core groups is above average. \textit{(D)} The distribution of roles with respect to KEGG pathway classification shows a markedly different contribution from each group.   \label{fig:figMET}}
\end{figure}

Our final example comes from metabolic networks, an area where identifying functional modules is crucial~\cite{Guimera:2008p24}. These networks have been analyzed using methods for undirected graphs, thus ignoring the inherent directionality of metabolite transformation in cellular pathways.  Figure~\ref{fig:figMET} shows our results for the largest connected component of the widely studied metabolic network of E.\ coli developed by Ma and Zeng \cite{Ma2003} from the Kyoto Encyclopedia of Genes and Genomes (KEGG)~\cite{kanehisa2000}.
Our results reveal the existence of a core, semi-peripheral and peripheral organization, a structure that is common among metabolic networks of many species and has been hypothesized previously~\cite{Holme:2005p4080}, but with a finer, more nuanced substructure. The network divides naturally into six significant groups including two types of input nodes, two types of cores, a set of intermediates and one group of outputs (Fig.~\ref{fig:figMET}B). We have used the extensive biological and functional characterization of metabolites in KEGG to examine the significance of these groups.   
Figure~\ref{fig:figMET}C shows that the metabolites in the core groups, and specifically those in Core 2, have a high metabolic importance, measured as the relative participation in different pathways. Hence these metabolites can be seen as forming the reservoir of cellular building blocks that are key to the function and interconnection of pathways in the cell.  In addition, we have characterized the KEGG pathway types in terms of our roles. Figure~\ref{fig:figMET}D shows, for instance, that central pathways such as carbohydrate and energy metabolism have an over-representation of core groups while, on the other hand, core groups are not involved in signaling pathways, which are dominated by a direct flow from input through intermediates to outputs. Unsurprisingly, biosynthetic internal pathways contain no inputs or outputs as they are used to generate intermediate and core metabolites. The detailed analysis of this functional classification will be presented elsewhere.

We have introduced here a conceptual basis for the grouping of nodes in directed networks based upon their role in the network, as established by the patterns of incoming and outgoing flows of all lengths. Our measure can be computed by taking successive powers of the adjacency matrix and convergence is ensured naturally within our definition. This measure formalizes and combines concepts present in the social network literature (e.g., structural equivalence) with ideas of similarity drawn from computer science.  In fact, one can show that the similarity matrix $Y$ can also be calculated iteratively based upon node similarity by computing the normalized sum of the convergent terms of:
\begin{eqnarray}
Y^{\mathrm{out}}_{n+1}&=&A \left (J+\left(\frac{\alpha}{\lambda_1}\right)^2 Y^{out}_n \right)A^T  \nonumber\\
Y^{\mathrm{in}}_{n+1}&=&A^T\left (J+\left(\frac{\alpha}{\lambda_1}\right)^2  Y^{in}_n \right )A, \nonumber
\end{eqnarray}
where $J$ is the matrix of ones and $Y_0$ is the matrix of zeros.
This algorithmic formulation allows for simplified updated computations in a format equivalent, yet functionally distinct, to other methods~\cite{Blondel:2004p2427, Leicht:2005}. In summary, our approach provides an alternative method to community detection algorithms for the simplification and abstraction of complex networks where directionality and flow transfer (rather than density of connections) is the fundamental ingredient to the description of the system. Our application to examples from a variety of fields highlights the applicability of such ideas across disciplines.

KC is supported by the Wellcome Trust.

\end{document}